# Magnetic Silicon Fullerene

Jing Wang,[a,b,c] Ying Liu*[a,b] and You-Cheng Li [a]

A metal-encapsulating silicon fullerene, Eu@$Si_{20}$, has been predicted by density functional theory to be by far the most stable fullerene-like silicon structure. The Eu@$Si_{20}$ structure is a dodecahedron with $D_{2h}$ symmetry in which the europium atom occupies the center site. The calculated results show that the europium atom has a large magnetic moment of nearly 7.0 Bohr magnetons. In addition, it was found that a stable "pearl necklace" nanowire, constructed by concatenating a series of Eu@$Si_{20}$ units, with the central europium atom retains the high spin moment. The magnetic structure of the nanowire indicates potential applications in the fields of spintronics and high-density magnetic storage.

## Introduction

Silicon, as a vital material for the vast semiconductor industry and one of the most studied elements in all of science, already has wide applications in computer chips, microelectronic devices, and new superconducting compounds. However, pure silicon cage clusters are unstable due to the lack of $sp^2$ bonding in silicon.[1] As carbon's neighbor in the same group of the periodic table, it might be hoped that it would also form structures analogous to carbon fullerenes. It is now thought possible that silicon cages with encapsulated metal atoms may lead to applications that could match or even exceed those expected for carbon fullerenes.[2] Researchers have been working to find some stable carbon-like silicon fullerenes. Several significant experimental works[2-4] have reported the formation of a series of metal-containing silicon clusters. By choosing an appropriate metal atom, the properties of these clusters can be tuned. On the theoretical side, there have been many first-principles investigations of silicon clusters with 8-20 atoms encapsulating 3$d$, 4$d$ and 5$d$ metal atoms.[5-11]

It is known that the smallest carbon fullerene cage is $C_{20}$ with a dodecahedral structure.[12,13] For the case of Si, many studies have shown that the ground state of pure $Si_{20}$ has a prolate structure based on stacking $Si_{10}$ tetracapped trigonal prism units.[14-17] Using first-principles calculations, Sun *et al.*[18] obtained distorted M@$Si_{20}$ cages doped with atoms such as Ba, Sr, Ca, Zr etc. The relatively small endohedral doping energies that were found are unlikely, however, to stabilize $Si_{20}$ fullerene. *Ab initio* electronic structure calculations have suggested that thorium should form a nonmagnetic neutral Th@$Si_{20}$ fullerene with icosahedral symmetry and that Th may be the only element that can stabilize the dodecahedral fullerene of $Si_{20}$.[19] Very recently, studies of the photoelectron spectra of $EuSi_n$ cluster anions ($3 \leq n \leq 17$) have shown that $EuSi_{12}^-$ is the smallest fully endohedral europium-silicon cluster.[20] By using the density functional approach, Wang *et al.*[21] calculated stabilities and electronic properties of novel transition bimetallic atoms encapsulated in a naphthalene-like $Si_{20}$ prismatic cage. So far the work on metal-encapsulated silicon clusters has focused on non-magnetic ones and very recently, Reveles *et al.*,[22] using a first-principles approach, reported a magnetic superatom, $VCs_8$, with large magnetic moments about 5 $\mu_B$. In the present work, it is shown that encapsulation of a europium atom into $Si_{20}$ yields by far the most stable magnetic M@$Si_{20}$ fullerene and high spin magnetic moment within the framework of density functional theory (DFT). Furthermore, a stable "pearl necklace" nanowire constructed of Eu@$Si_{20}$ subunits was found to have high spin moment. It may have important applications in the fields of spintronics and high-density magnetic storage, such as a magnetic field controlled nanowire which can produce and transport the spin-polarized current.

## Theoretical Methods

In the course of the geometry optimizations and the total-energy calculations, the exchange-correlation interaction was treated within the GGA using two different exchange-correlation functionals, the Perdew-Wang (PW91)[23] and the Becke exchange plus Lee-Yang-Parr correlation (BLYP)[24], to reduce uncertainties associated with the numerical procedures, as well as to give a comparison of different functionals. A double-numerical polarized (DNP) basis set[25] with unrestricted spin was chosen to carry out the electronic structure calculation. The optimizations were carried out without symmetry restrictions. All computations were carried out in the DMol$^3$ program package.[26]

At first, optimizations were performed for metal-encapsulated $Si_{20}$ fullerene cages. To seek out a suitable embedded atom, a broader search were made through the elements (M) with relatively large atomic radius in the periodic table, such as the late alkali metals, the late haloid elements, the 3$d$ and 4$d$ transition metals, and the lanthanide series and so on. At the same time, their stabilities were further test. By comparison with the M@$Si_{20}$ cages, a number of initial structures of $MSi_{20}$ were also investigated by substituting one of the silicon atoms of the accepted lowest-energy $Si_{21}$[27] with the M atom and the results of the $MSi_{20}$ clusters with some typical embedded elements (Rb, Cs, La, Eu, and Gd) are listed in Table 1. Following the optimization, the stability of Eu@Si$_{20}$ was further checked by (*a*) vibrational frequency analysis and (*b*) molecular dynamics (MD) simulation. Here, an *ab initio* MD in the constant-NVT ensemble with a Massive Nosé-Hoover thermostat was imposed on stable Eu@Si$_{20}$ cage at the PW91 level. The simulation run was with a time step of 1.0 femtosecond and 10000 dynamics steps and the average temperature was fixed at 1000 K.

## Results and Discussions

Europium, the rare earth element with the atomic number of 63, has the ground state valence-electron configuration of $4f^7 6s^2$. The large valence shell orbital radii, large atomic magnetic moment and half-filled 4*f* shell orbital of Eu atom indicates that it's an ideal candidate as "guest" to produce a carbon-like silicon fullerene.

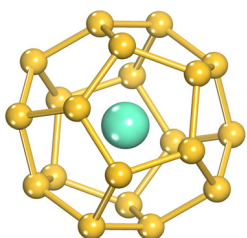

**Fig.1** The stable Eu@Si$_{20}$ fullerene. Large ball: Eu atom; small ball: Si atom.

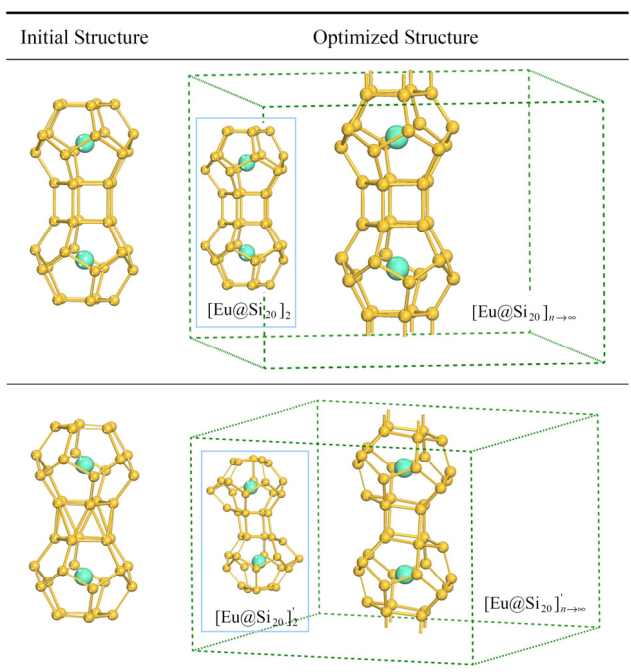

**Fig.2** The initial and optimized structures of [Eu@Si$_{20}$]$_2$ and [Eu@Si$_{20}$]$_2'$ dimers and the unit cells of the two optimized pear necklace nanowires, [Eu@Si$_{20}$]$_{n\to\infty}$ and [Eu@Si$_{20}$]$_{n\to\infty}'$. Large ball: Eu atom; small ball: Si atom.

From the values of Table 1, it can be seen that the optimized Rb@Si$_{20}$ and Cs@Si$_{20}$ cages are obviously the high-lying isomers with relatively large energy differences to the lowest-energy RbSi$_{20}$ and CsSi$_{20}$ structures. For CsSi$_{20}$, at the PW91 level, the lowest-energy structure is No. 14 and the energy difference to that of the Cs@Si$_{20}$ cage reaches 5.68 eV. For La, Eu, and Gd, the optimizations of the *M*-encapsulated $I_h$-Si$_{20}$ cages obtained the lowest-energy structures. For Eu, the optimization of the cage structure leads to a stable Eu@Si$_{20}$ fullerene cage with $D_{2h}$ symmetry. The $D_{2h}$-Eu@Si$_{20}$ cage is energetically favored by 0.79 eV lower in energy than that of the second low-lying isomer (No. 18) at the PW91 level. All these results indicate that the rare earth element (R) can stabilize the Si$_{20}$ cage in the neutral state by forming a R@Si$_{20}$ fullerene cage. Remarkably, the europium atom has a half-filled *f*-orbit, and Eu@Si$_{20}$ cage keeps the large magnetic moment.

Figure 1 gives the structure of the stable $D_{2h}$-Eu@Si$_{20}$ fullerene cage, showing that it is a Si$_{20}$ dodecahedron with complete encapsulation of the europium atom. Then the result of the vibrational frequency analysis at the PW91 level further confirms the structural stablity of $D_{2h}$-Eu@Si$_{20}$ with no imaginary frequency[29]. Moreover, the validation of kinetic stability by an *ab initio* MD indicates that the bond structure of Eu@Si$_{20}$ keeps the same topological structure of $D_{2h}$-Eu@Si$_{20}$[30].

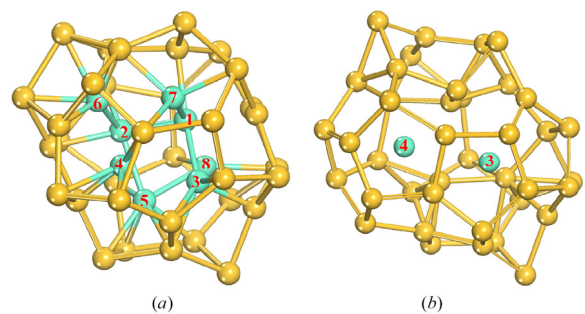

**Fig. 3** (*a*) The configuration of Si$_{42}$[28] with the eight internal sites highlighted. By substituting two Eu atoms for two of the eight highlighted Si atoms, 28 initial structures of Eu$_2$@Si$_{40}$ were obtained. (*b*) The lowest-energy isomer of Eu$_2$@Si$_{40}$ among the 28 isomers. The numbers represent different inner sites.

Based on the analysis of the electronic structure of the stable Eu@Si$_{20}$ fullerene cage, a large spin magnetic moment of nearly 7.0 $\mu_B$ was found for the central europium atom. On the other hand, the HOMO-LUMO gap of this structure is only about 0.2 eV indicating relatively large chemical reactivity.

Furthermore, the stability and electronic structure of the dimers constructed by concatenating two Eu@Si$_{20}$ subunits were investigated. Two kinds of dimers, [Eu@Si$_{20}$]$_2$ and [Eu@Si$_{20}$]$_2'$, were constructed by concatenating two Eu@Si$_{20}$ subunits without and with a rotation, as shown in Fig. 2. Optimizitions indicate the existence of a stable [Eu@Si$_{20}$]$_2$ dimer. It can be seen that the stable [Eu@Si$_{20}$]$_2$ dimer keeps the Eu@Si$_{20}$ subunit intact, while distortions occur for [Eu@Si$_{20}$]$_2'$ dimer. At the same time, calculations were imposed on Eu$_2$@Si$_{40}$ clusters for stability comparison. By substituting two Eu atoms for two of the eight internal Si atoms of a lowest-energy structure of Si$_{42}$[28], twenty eight initial structures of Eu$_2$@Si$_{40}$ were obtained. The configuration of Si$_{42}$ is shown in Fig. 3 with the eight internal silicon atoms highlighted. The calculated binding energies of $Eu_2@Si_{40}$ are listed in Table 3. It can be seen that No.(3,4) has the lowest binding energy among the 28 $Eu_2@Si_{40}$ clusters. Compared with the stable $[Eu@Si_{20}]_2$ dimer, No.(3,4) is only 0.19 eV higher in energy than that of $[Eu@Si_{20}]_2$, but more stable than $[Eu@Si_{20}]'_2$ (-160.21 eV at the PW91 level) with 0.77 eV energy difference.

For the lowest-energy $[Eu@Si_{20}]_2$, the central Eu atoms keep the large spin magnetic moments of nearly 7.0 $\mu_B$, as listed in Table 2. It is interesting to note that the spin moments of $[Eu@Si_{20}]_2$ and $[Eu@Si_{20}]'_2$ dimers display different characters. For the lowest $[Eu@Si_{20}]_2$ dimer, the spin moments of the two central Eu atoms have the same direction, while for $[Eu@Si_{20}]'_2$ they are opposite, one spin up and the other spin down.

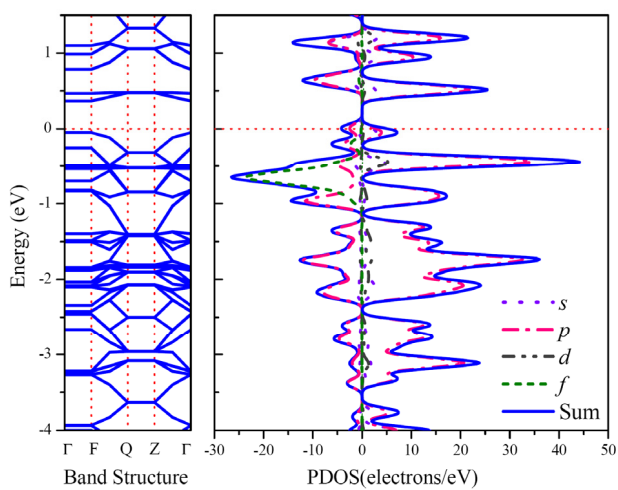

**Fig. 4** The band structure and the partial density of states (PDOS) including the spin up and spin down for $[Eu@Si_{20}]_{n\to\infty}$ nanowire.

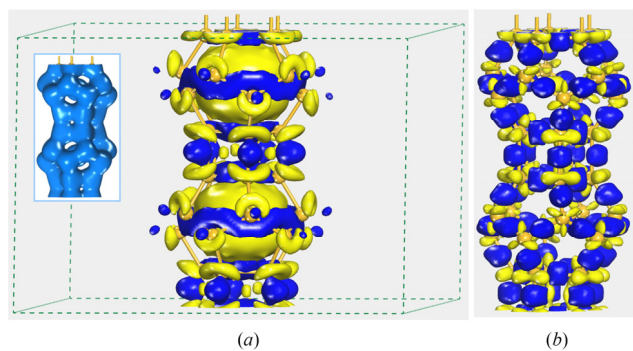

**Fig.5** (*a*) The spin density and (*b*) the deformation electron density of $[Eu@Si_{20}]_{n\to\infty}$. The insert of (*a*) is the corresponding total electron density. The iso-values are $7.0\times10^{-4}$ and 0.18 e/Å$^3$ for spin density and total electron density, and 0.035 e/Å$^3$ for deformation electron density.

Then the quasi-one-dimensional nanowires constructed of a series of stable $Eu@Si_{20}$ subunits were investigated. Optimizations indicate the existence of a stable pearl necklace nanowire, $[Eu@Si_{20}]_{n\to\infty}$, as shown in Fig. 2, the unit cell of the infinite $Eu@Si_{20}$ chains. It can be seen that $[Eu@Si_{20}]_{n\to\infty}$ linear structure obtained by concatenating two $Eu@Si_{20}$ subunits without a rotation keeps the stable $Eu@Si_{20}$ subunit intact. Close examination of the band structure and the partial density of state (PDOS) of $[Eu@Si_{20}]_{n\to\infty}$ nanowire, as shown in Fig. 4, indicate that it appears to have some of the properties of a direct-gap semiconductor with 0.42 eV energy gap at the $\Gamma$ point.

The spin density and the deformation electron density of $[Eu@Si_{20}]_{n\to\infty}$, as well as the total electron density are shown in Fig. 5. It can be seen from the figure that the Si-Si bonding coexists the $sp^3$-like and $sp^2$-like hybridizations. For each Si atom around the neck, it resembles the $sp^3$ and lobes point towards the four neighboring Si atoms. For the equatorial Si atoms, each has three neighbors. Three lobes from each Si atom form $\sigma$ bonds with three neighboring Si atoms, which leads to the formation of $sp^2$-like hybridization. It is noteworthy that for the pearl necklace nanowire the spin moments of the Eu atoms remain as large as that in the ground-state $D_{2h}$-$Eu@Si_{20}$ fullerene, the same as $[Eu@Si_{20}]_2$ dimer. In addition, the spin magnetic moment of the central Eu atom, for each $Eu@Si_{20}$ subunit of $[Eu@Si_{20}]_{n\to\infty}$, has the same direction, which may have some potential applications in spin-polarized-current switch controlled by external magnetic field.

**Table 1.** The binding energy ($E_b$) in unit of eV for $MSi_{20}$ ($M$=Rb, Cs, La, Eu, and Gd) obtained with the PW91/BLYP exchange correlation. No. 1-21 represent different initials coming from substitution of $M$ for a Si of the accepted lowest-energy $Si_{21}$, and cage for the $M$-encapsulated $Si_{20}$ fullerene.

| | $E_b$ (PW91/BLYP) | | | | |
|---|---|---|---|---|---|
| | $RbSi_{20}$ | $CsSi_{20}$ | $LaSi_{20}$ | $EuSi_{20}$ | $GdSi_{20}$ |
| **Cage** | **-72.22/-62.47** | **-71.01/-60.90** | **-82.08/-71.33** | **-78.34/-67.76** | **-81.61/-71.00** |
| No.1 | -76.29/-65.84 | -76.45/-66.15 | -79.02/-68.47 | -76.30/-65.94 | -78.56/-68.02 |
| No.2 | -76.14/-66.02 | -76.29/-66.16 | -79.69/-68.91 | -77.00/-66.53 | -79.08/-68.40 |
| No.3 | -75.88/-65.40 | -76.05/-65.58 | -79.06/-67.94 | -76.78/-65.89 | -78.38/-67.50 |
| No.4 | -75.04/-65.13 | -75.16/-65.24 | -78.52/-68.00 | -75.11/-64.96 | -78.11/-67.61 |
| No.5 | -76.02/-65.88 | -76.20/-66.01 | -79.39/-68.66 | -76.95/-66.43 | -78.80/-68.29 |
| No.6 | -75.03/-65.12 | -75.15/-65.23 | -79.55/-68.77 | -76.54/-66.01 | -78.83/-68.11 |
| No.7 | -76.14/-65.96 | -76.30/-66.10 | -78.73/-68.01 | -76.44/-65.91 | -78.14/-67.65 |
| No.8 | -75.67/-65.49 | -75.83/-65.64 | -79.49/-68.53 | -76.87/-66.19 | -78.86/-68.03 |
| No.9 | -75.86/-65.61 | -76.07/-65.66 | -79.54/-68.76 | -76.65/-65.92 | -78.35/-67.24 |
| No.10 | -75.75/-65.53 | -75.93/-65.69 | -79.24/-68.31 | -76.92/-66.22 | -78.59/-67.74 |
| No.11 | -76.11/-65.71 | -76.29/-65.87 | -79.77/-68.83 | -77.04/-66.37 | -79.21/-68.33 |
| No.12 | -75.64/-65.48 | -75.83/-65.65 | -78.65/-67.85 | -75.98/-65.49 | -78.10/-67.36 |
| No.13 | -76.08/-65.96 | -76.28/-66.15 | **-79.96/-68.92** | -77.23/-66.59 | **-79.28**/-68.37 |
| No.14 | -76.25/-66.08 | **-76.69**/-66.31 | -79.21/-68.25 | -76.46/-65.76 | -78.59/-67.70 |
| No.15 | **-76.34/-66.15** | -76.52/**-66.32** | -79.87/-69.00 | -77.18/-66.54 | -79.26/**-68.45** |
| No.16 | -75.64/-65.53 | -74.56/-64.50 | -78.72/-67.91 | -75.76/-65.21 | -77.99/-67.29 |
| No.17 | -76.14/-66.01 | -76.34/-66.20 | -79.53/-68.75 | -77.18/-66.54 | -78.78/-68.08 |
| No.18 | -76.35/-66.04 | -76.55/-66.22 | -79.44/-68.63 | **-77.55/-66.88** | -79.04/-68.32 |
| No.19 | -76.13/-65.97 | -76.33/-66.16 | -79.61/-68.81 | -77.24/-66.64 | -78.96/-68.20 |
| No.20 | -76.12/-65.94 | -76.32/-66.13 | -79.64/-68.84 | -77.28/-66.67 | -78.98/-68.36 |
| No.21 | -76.14/-65.77 | -76.34/-65.96 | -79.75/-68.88 | -77.05/-66.39 | -79.17/-68.35 |

**Table 2.** Calculated results for the stable $Eu@Si_{20}$ fullerene, the $[Eu@Si_{20}]_2$ dimer, and $[Eu@Si_{20}]_{n\to\infty}$ nanowire at the PW91/BLYP level. The columns give data for the binding energy ($E_b$), the HOMO-LUMO gap (Gap), the spin moment ($\mu_s$) and atomic charge ($Q$) of the Eu atom, the total spin moment ($\mu_T$) and unpaired electrons ($q$) of each molecular system or per unit cell for nanowire.

| Type | $E_b$(eV) | Gap(eV) | $\mu_s(\mu_B)$ | $Q$ (e) | $\mu_T(\mu_B)$ | $q$ |
|---|---|---|---|---|---|---|
| Eu@Si$_{20}$ | -78.34/ -67.76 | 0.31/0.21 | 7.02/6.88 | -1.51/ -1.00 | 5.00/ 9.00 | 5/3 |
| [Eu@Si$_{20}$]$_2$ | -161.27/ -138.96 | 0.29/0.19 | 6.90/6.96 | -1.33/ -0.92 | 11.99/ 12.00 | 12/19 |
| [Eu@Si$_{20}$]$_{n\to\infty}$ | -165.78/ -142.25 | -- | 7.01/6.96 | -0.97/ -0.54 | 13.99/ 13.99 | 13/13 |

**Table 3.** The binding energy ($E_b$) of Eu$_2$@Si$_{40}$ obtained at the PW91 exchange correlation. The ($n$, $m$) with $n$, $m$ = 1-8 ($n<m$) represents different initials coming from substitutions of two Eu atoms for two of the eight internal Si atoms of a lowest-energy structure of Si$_{42}$[28] and also see Fig. 3 for $n$ and $m$.

| No. | $E_b$(eV) | No. | $E_b$(eV) | No. | $E_b$(eV) | No. | $E_b$(eV) |
|---|---|---|---|---|---|---|---|
| (1,2) | -159.80 | (2,3) | -159.20 | (3,5) | -160.37 | (4,8) | -158.79 |
| (1,3) | -158.99 | (2,4) | -160.25 | (3,6) | -160.28 | (5,6) | -159.74 |
| (1,4) | -159.04 | (2,5) | -158.49 | (3,7) | -159.13 | (5,7) | -158.58 |
| (1,5) | -159.40 | (2,6) | -159.94 | (3,8) | -159.35 | (5,8) | -157.99 |
| (1,6) | -158.47 | (2,7) | -159.22 | (4,5) | -159.35 | (6,7) | -159.68 |
| (1,7) | -160.11 | (2,8) | -158.98 | (4,6) | -158.37 | (6,8) | -160.19 |
| (1,8) | -160.83 | **(3,4)** | **-161.08** | (4,7) | -158.19 | (7,8) | -159.36 |

## Conclusions

In summary, DFT-GGA calculations suggest that the rare earth atoms, La, Eu, and Gd, can stabilize the Si$_{20}$ fullerene cage, and for $D_{2h}$-Eu@Si$_{20}$, the central europium atom keeps the high spin magnetic moment. The $D_{2h}$-Eu@Si$_{20}$ fullerene can be used as a repeat unit to construct the stable quasi-one-dimensional pearl necklace nanowire. Subsequently, the [Eu@Si$_{20}$]$_{n\to\infty}$ is a semiconductor and the Eu atom of the nanowire retains its high spin magnetic moment. Given the properties of these structures, there may be significant potential for exploiting novel materials based on Eu@Si$_{20}$ in spintronics devices or for high-density magnetic storage.

## Acknowledgements

We would like to thank Professor S. S. Li, N. X. Chen and J. Shen for their helpful discussions and N. E. Davison for his help with the language. This work was supported by the National Basic Research Program of China (Grant No. 2006CB605101), the National Natural Science Foundation of China (Grant No.10874039) and the Natural Science Foundation of Hebei Province (No.A2008000134).

## Notes and references

*a* Department of Physics, and Hebei Advanced Thin Film Laboratory, Hebei Normal University, Shijiazhuang 050016, Hebei, China. Fax: +86-311-86268314; Tel: +86-311-86268649; E-mail: yliu@hebtu.edu.cn.
*b* National Key Laboratory for Materials Simulation and Design, Beijing 100083, China.
*c* State Key Laboratory for Superlattices and Microstructures, Institute of Semiconductors, Chinese Academy of Sciences, Beijing 100083, China.
*Corresponding author Email: yliu@hebtu.edu.cn.

1  K. M. Ho, A. A. Shvartsburg, B. Pan, Z. Y. Lu, C. Z. Wang, J. G. Wacker, J. L. Fye and M. F.Jarrold, *Nature (London)*, 1998, **392**, 582.
2  H. Hiura, T. Miyazaki and T. Kanayama, *Phys. Rev. Lett.*, 2001, **86**, 1733; http://www.aip.org/enews/physnews/2001/split/527-1.html.
3  S. M. Beck, *J. Chem. Phys.*, 1987, **87**, 4233; S. M. Beck, *J. Chem. Phys.*, 1989, **90**, 6306.
4  K. Koyasu, M. Akutsu, M. Mitsui and A. Nakajima, *J. Am. Chem. Soc.*, 2005, **127**, 4998.
5  V. Kumar and Y. Kawazoe, *Phys. Rev. Lett.*, 2001, **87**, 0455034.
6  J. Lu and S. Nagase, *Phys. Rev. Lett.*, 2003, **90**, 115506.
7  J. U. Reveles and S. N. Khanna, *Phys. Rev. B*, 2006, **74**, 035435.
8  M. B. Torres, E. M. Fernández and L. C. Balbás, *Phys. Rev. B*, 2007, **75**, 205425.
9  Q. Peng and J. Shen, *J. Chem. Phys.*, 2008, **128**, 084711.
10  Q. Peng, J. Shen and N. X. Chen, *J. Chem. Phys.*, 2008, **129**, 034704.
11  J. Wang, Q. M. Ma, Z. Xie, Y. Liu and Y. C. Li, *Phys. Rev. B*, 2007, **76**, 035406.
12  H. Prinzbach, A. Weiler, P. Landenberger, F. Wahl, J. Wrth, L. T. Scott, M. Gelmont, D. Olevano and B. Issendorff, *Nature (London)*, 2000, **407**, 60.
13  M. F. Jarrold, *Nature (London)*, 2000, **407**, 26.
14  A. A. Shvartsburg, B. Liu, Z. Lu, C. Z. Wang, M. F. Jarrold and K. M. Ho, *Phys. Rev. Lett.*, 1999, **83**, 2167.
15  L. Mitas, J. C. Grossman, I. Stich and J. Tobik, *Phys. Rev. Lett.*, 2000, **84**, 14792.
16  B. X. Li and P. L. Cao, *Phys. Rev. A*, 2000, **62**, 023201.
17  J. Wang, J. Zhao, F. Ding, W. Shen, H. Lee and G. H. Wang, *Solid State Commun.*, 2001, **117**, 593.
18  Q. Sun, Q. Wang, T. M. Briere, V. Kumar and Y. Kawazoe, *Phys. Rev. B*, 2002, **65**, 235417.
19  A. K. Singh, V. Kumar and Y. Kawazoe, *Phys. Rev. B*, 2005, **71**, 115429.
20  A. Grubisic, H. P. Wang, Y. J. Ko and K. H. Bowena, *J. Chem. Phys.*, 2008, **129**, 054302.
21  J. Wang and J. H. Liu, *J. Comput. Chem.*, 2009, **30**, 1103.
22  J. U. Reveles, P. A. Clayborne, A. C. Reber1, S. N. Khanna, K. Pradhan, P. Sen and M. R. Pederson, *Nature Chem.*, 2009, **1**, 310.
23  J. P. Perdew and Y. Wang, *Phys. Rev. B*, 1992, **45**, 13244.
24  A. D. Becke, *J. Chem. Phys.*, 1988, **88**, 2547; C. Lee, W. Yang and R. G. Parr, *Phys. Rev. B*, 1988, **37**, 785.
25  B. Delley, *J. Chem. Phys.*, 1990, **92**, 508.
26  B. Delley, *J. Chem. Phys.*, 2000, **113**, 7756.
27  S. Yoo and X. C. Zeng, *J. Chem. Phys.*, 2006, **124**, 054304.
28  The lowest-energy structure of Si$_{42}$ is derived from the Si$_{40}$-Si$_{60}$-PLA.rar at http://www-wales.ch.cam.ac.uk/~wales/CCD/Si.html, and the corresponding reference is R. L. Zhou, B. C. Pan, *Phys. Lett. A*, 2007, **368**, 396.
29  The frequency spectrum obtained at the PW91 level are shown as Supplementary information (See Part I).
30  An *ab initio* molecular dynamics was imposed on the $D_{2h}$-Eu@Si$_{20}$ cage. See Part II of Supplementary information for the curve of temperature *vs* dynamics steps.